# Etude expérimentale de l'alimentation d'un dispositif supraconducteur à courant continu

Lauro FERREIRA[1,2], Yasmine BAAZIZI[1,2], Simon MEUNIER[1,2], Tanguy PHULPIN[1,2], Richard BELJIO[1,2], Frédéric TRILLAUD[3], Tian-Yong GONG[1,2], Gustavo HENN[1,2], Loïc QUÉVAL[1,2]

[1] Université Paris-Saclay, CentraleSupélec, CNRS, GeePs, 91192, Gif-sur-Yvette, France. [2] Sorbonne Université, CNRS, GeePs, 75252, Paris, France. [3] Instituto de Ingeniería, Universidad Nacional Autonoma de Mexico, 04510 Mexico city, Mexique.

**RESUMÉ** - Bien qu'un supraconducteur ne présente aucune perte en courant continu, un système supraconducteur présente lui des pertes non négligeables, notamment à cause de son alimentation. Nous étudions ici deux systèmes d'alimentation différents. Le premier, qui est conventionnel, est constitué d'un transformateur et d'un pont de diodes fonctionnant à température ambiante ainsi que d'amenées de courant qui permettent le passage du courant du milieu à température ambiante au milieu cryogénique. Le second comprend un transformateur, dont l'enroulement secondaire est supraconducteur, associé à un pont de diodes fonctionnant à température cryogénique ce qui permet de se passer des amenées de courant. Nous comparons expérimentalement les performances des transformateurs conventionnel et supraconducteur ainsi que les performances d'un pont de diodes à température ambiante et cryogénique. Nos résultats indiquent que le prototype de transformateur supraconducteur développé présente une résistance des enroulements et une inductance de fuite ramenée au secondaire plus faibles que le transformateur conventionnel. En outre, nous avons constaté que seules certaines diodes sont adaptées au fonctionnement à température cryogénique. Enfin, le pont de diodes composé à partir de diodes adaptées présente des pertes réduites à température cryogénique. Ces travaux expérimentaux sont la première étape de la réalisation d'un système d'alimentation complet d'un dispositif supraconducteur.

*Mots-clés* — *Transformateur supraconducteur, Cryoconvertisseur, Supraconductivité, Electronique de puissance, Efficacité énergétique, Densité de puissance.*

## 1. INTRODUCTION

La supraconductivité offre un potentiel considérable pour la conception de dispositif supraconducteurs pour de multiples domaines d'application, notamment la spectrométrie de masse, la fusion nucléaire, la physique des particules ou encore la lévitation magnétique pour les véhicules maglev [1].

Bien qu'un supraconducteur (SC) ne présente aucune perte en courant continu, un système supraconducteur présente lui des pertes non négligeables. Prenons l'exemple du système conventionnel montré sur la Fig. 1a : il s'agit d'un dispositif supraconducteur à courant continu alimenté depuis une source de courant alternatif connectée en série avec un pont de diodes. Des pertes se manifestent dans chaque composant de la chaîne de conversion. Notons qu'une partie importante des pertes est liée aux amenées de courants, un composant qui permet le passage du courant du milieu à température ambiante au milieu cryogénique [2].

Nous étudions une option alternative à ce système, sans amenée de courant : un transformateur dont l'enroulement secondaire est supraconducteur associé à un pont de diodes opérant à température cryogénique (Fig. 1b). Dans cet article, nous rapportons les premiers résultats expérimentaux concernant le transformateur supraconducteur et le pont de diodes cryogénique.

La suite de cet article est divisée en trois sections. La section 2 discute les avantages et les inconvénients du système étudié par rapport au système conventionnel. Les sections 3 et 4 présentent respectivement le transformateur supraconducteur et le pont de diodes cryogénique.

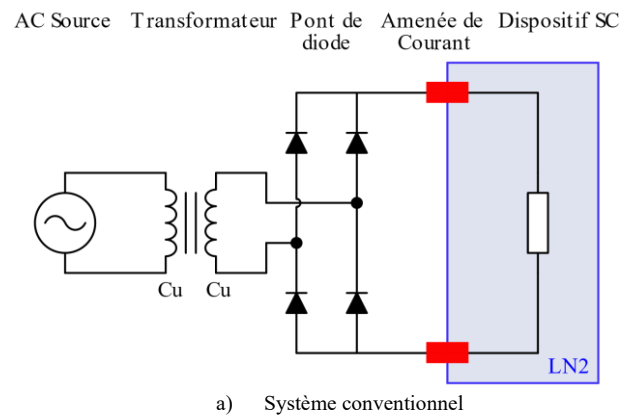

a) Système conventionnel

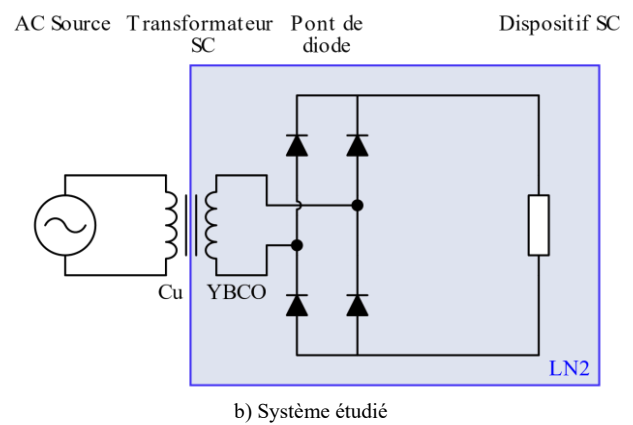

b) Système étudié

Fig. 1 - Systèmes d'alimentation d'un dispositif supraconducteur à courant continu.

## 2. ALIMENTATION D'UN DISPOSITIF SUPRACONDUCTEUR

Dans le système conventionnel Fig. 1a), le dispositif supraconducteur est connecté à la source monophasée en courant alternatif par l'intermédiaire d'un transformateur monophasé dont les enroulements primaires et secondaires sont résistifs, et d'un pont de diodes fonctionnant à température ambiante, ainsi que de deux amenées de courant.

Dans le système étudié (Fig. 1b), le dispositif supraconducteur est connecté à la source par l'intermédiaire d'un transformateur monophasé dont l'enroulement primaire est résistif et l'enroulement secondaire est supraconducteur, et d'un pont de diodes fonctionnant à température cryogénique. Ce système permet de se passer des amenées de courant, en transférant l'énergie par induction électromagnétique à travers la paroi du cryostat.

Si l'utilisation d'un transformateur supraconducteur peut permettre un gain d'efficacité, la compacité et la légèreté du dispositif sont souvent citées comme des points clés [3]. De même, l'utilisation d'un pont de diodes à température cryogénique permet de bénéficier de l'amélioration des performances de certains composants semi-conducteurs à ces températures. Mais l'avantage principal est que le volume et le poids du dissipateur thermique peuvent être considérablement réduit [4]. La solution étudiée pourrait donc être une solution prometteuse pour les applications soumises à des contraintes de volume et de masse, telles que l'énergie éolienne, le transport aérien ou le transport naval.

Inversement, l'utilisation d'un plus grand nombre de composants à une température cryogénique augmente la charge thermique du système de refroidissement. Le facteur d'efficacité de celui-ci étant relativement faible, on peut s'attendre à une consommation d'énergie supplémentaire importante, ce qui peut aboutir à une solution non compétitive. Cette considération a un poids moindre dans le cas où la puissance de refroidissement est disponible gratuitement, comme dans le cas du futur avion à hydrogène, où l'hydrogène serait stocké à une température cryogénique avant d'être réchauffé pour être utilisé comme carburant [5].

## 3. TRANSFORMATEUR

### 3.1. Vue d'ensemble

Un prototype de transformateur supraconducteur, de taille réduite, a été réalisé. Il s'agissait d'une première au laboratoire GeePs : le but était donc d'identifier les difficultés avant de réaliser un prototype plus ambitieux. Bien que cette topologie ne soit pas optimale, nous adoptons un noyau magnétique de type C-C qui permet un couplage magnétique simple à travers la paroi du cryostat (Fig. 2). Cette topologie est utilisée pour les deux systèmes, avec le même entrefer, afin de faciliter la comparaison.

▪ Système conventionnel - Les deux enroulements du transformateur sont résistifs et fonctionnent à température ambiante. Pour notre prototype (Fig. 2a), l'enroulement primaire est bobiné de 56 tours avec un conducteur en cuivre de section 1 mm$^2$. L'enroulement secondaire est bobiné de 14 tours avec un conducteur en cuivre de section 3.5 mm$^2$. Les enroulements sont bobinés sur des carcasses en plastique, chacune enfilée sur un noyau magnétique de type C.

▪ Système étudié - L'enroulement primaire du transformateur est résistif et fonctionne à température ambiante. L'enroulement secondaire est supraconducteur et fonctionne à température cryogénique. L'un des noyaux magnétiques opère également à température cryogénique. Pour notre prototype (Fig. 2b), nous utilisons le même enroulement primaire que pour le système conventionnel. L'enroulement secondaire est lui bobiné de 14 tours avec un ruban supraconducteur 2G YBCO (SuNAM SCN04150-201013-03). Il est refroidi à 77 K dans un bain ouvert d'azote liquide.

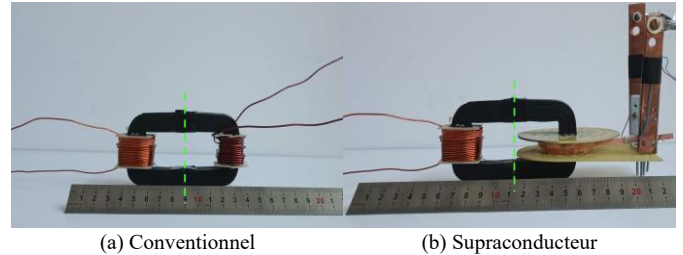

(a) Conventionnel      (b) Supraconducteur

Fig. 2 - Prototypes de transformateur. Les pointillés montrent la position de l'entrefer.

Un bobinage supraconducteur en double pancake [6] a été retenu afin de minimiser le nombre de jonction et de déporter les connecteurs en périphérie de bobine pour pouvoir facilement connecter plusieurs bobines en séries par la suite. Une bobineuse permettant de réaliser ce type de bobinage a été développé. À la suite du bobinage, la caractéristique V-I de la bobine a été mesurée en présence du noyau magnétique à 77 K (Fig. 3). La caractéristique indique que le double pancake est bien supraconducteur après bobinage. Cependant, le courant critique du double pancake est de seulement 33 A, alors que le courant critique du ruban est d'environ 250 A. Ceci peut s'expliquer par un certain nombre de facteurs qu'il conviendra d'étudier : faible rayon de courbure (15 mm), présence du noyau magnétique, manipulations du ruban lors du bobinage, connecteurs, etc.

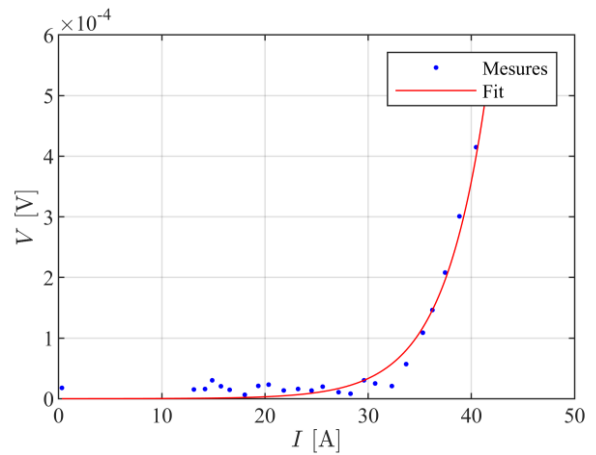

Fig. 3 - Caractéristique tension-courant mesurée du double pancake supraconducteur en présence du noyau magnétique.

### 3.2. Résultats expérimentaux

Pour les tests, le primaire et le secondaire sont séparés par la paroi d'un cryostat non magnétique d'une épaisseur de 4 mm (Fig. 4). Le primaire est connecté à une source de tension alternative. Le secondaire est connecté à une charge résistive. Les tensions et courants du primaire ($v_1, i_1$) et du secondaire ($v_2, i_2$) des deux transformateurs, pour la même source et la même charge, sont tracés sur les Figs. Fig. 5 - Fig. 6.

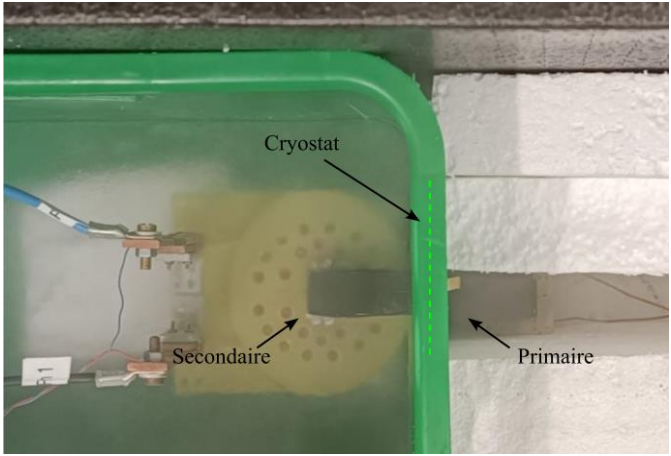

Fig. 4 - Test du transformateur supraconducteur à température cryogénique.

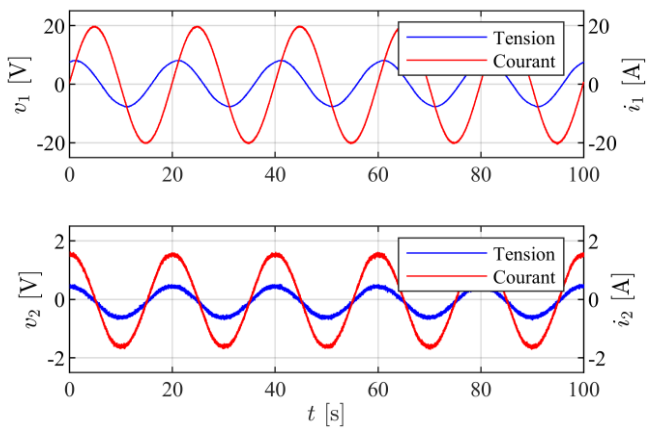

Fig. 5 - Transformateur conventionnel : tension et courant du primaire et du secondaire.

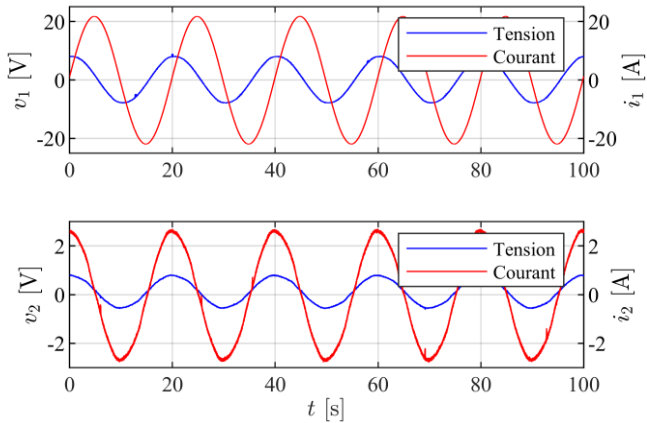

Fig. 6 - Transformateur supraconducteur : tension et courant du primaire et du secondaire.

Les paramètres du modèle équivalent du transformateur monophasé (Fig. 7) sont résumés dans le Tab. 1. Les rendements mesurés sur ces premiers prototypes sont très faibles : le but était de tester les capacités de réalisation d'un bobinage supraconducteur, aucun effort n'a été fait au niveau conception. De plus la topologie avec noyau magnétique de type C-C et un entrefer n'est pas la plus adaptée pour la réalisation d'un transformateur. Quoi qu'il en soit, ces premiers prototypes permettent de valider certains points.

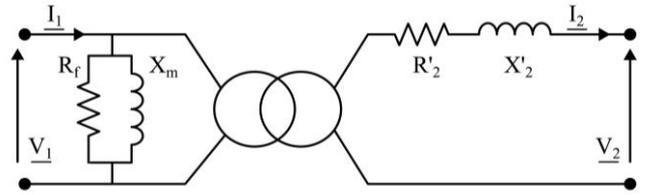

Fig. 7 - Modèle équivalent d'un transformateur monophasé.

Tab. 1 - Paramètres des transformateurs.

|  | $V_2/V_1$ | $\eta_t$ [%] | $R_f$ [Ω] | $X_m$ [Ω] | $R'_2$ [Ω] | $X'_2$ [Ω] |
|---|---|---|---|---|---|---|
| Conv. | 0,11 | 1,5 | 1,13 | 0,42 | 3,76 | 4,24 |
| Supra. | 0,09 | 25,8 | 1,74 | 0,38 | 0,05 | 0,16 |

*Note :* L'efficacité ne tient pas compte de l'efficacité du système de refroidissement ($\eta_t = P_2/P_1$).

Premièrement, aucune différence majeure n'est observée lorsqu'une partie du noyau magnétique fonctionne à température cryogénique (*i.e.* les valeurs de $R_f$ et $X_m$ sont similaires à température ambiante et cryogénique). Deuxièmement, la résistance des enroulements du transformateur supraconducteur est plus faible que celle du transformateur conventionnel, comme attendu. Troisièmement, l'inductance de fuite ramenée au secondaire est plus faible dans le cas du transformateur supraconducteur ; cela est peut-être dû à la disposition en double pancake du secondaire supraconducteur.

## 4. PONT DE DIODES

### 4.1. Selection d'une diode adaptée

Différentes technologies de semi-conducteurs présentent des comportements différents à température cryogénique [7]. Afin de choisir une diode de puissance adaptée, de nombreuses références ont été testées. Nous rapportons ici uniquement les caractéristiques courant-tension (I-V) mesurées à 77 K de deux diodes différentes.

La première est la BYT60-400, une diode de redressement à récupération rapide fabriquée par SGS-Thomson Microelectronics. Elle est en silicium et se présente sous la forme d'un boîtier DO-5. La deuxième diode est la VS-HFA220FA120, une diode de redressement à récupération ultra-rapide fabriquée par Vishay Semiconductors. Elle est aussi en silicium mais se présente dans un boîtier SOT-227.

Les résultats expérimentaux sont présentés sur la Fig. 8 et les paramètres des diodes sont résumés dans le Tab. 2. La courbe I-V de la diode BYT60-400 se translate vers la droite à température cryogénique. Par conséquent, sa tension de seuil augmente. Cela indique que les pertes par conduction de cette diode augmentent lorsqu'elle fonctionne à température cryogénique, et que par conséquent cette diode n'est pas adaptée à un fonctionnement à basse température.

La courbe I-V de la diode VS-HFA220FA120 se déplace vers la gauche et sa pente augmente à température cryogénique. Ainsi, sa résistance et sa tension de seuil diminuent à 77 K. Cette diode présente moins de pertes par conduction lorsqu'elle fonctionne à basse température. Cette diode est donc adaptée pour une utilisation à température cryogénique et est retenue pour la construction du pont de diodes monophasés.

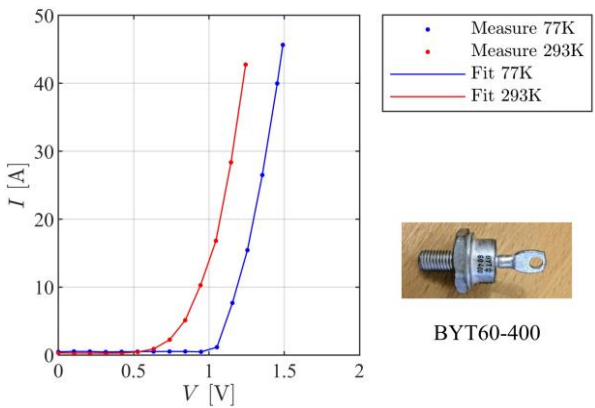

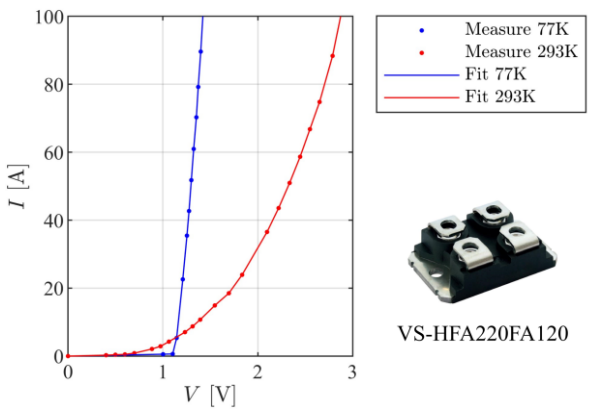

Fig. 8 - Caractéristiques courant-tension mesurées de deux diodes de puissance différentes.

Tab. 2 - Paramètres des diodes mesurés à température ambiante et à température cryogénique.

| Symbole | Paramètre | Temp. | BYT60-400 | VS-HFA 220FA120 |
|---|---|---|---|---|
| $R_{on}$ | Resistance passante | 77 K | 8,60 mΩ | 2,67 mΩ |
|  |  | 293 K | 10,27 mΩ | 14,40 mΩ |
| $V_{t0}$ | Tension de seuil | 77 K | 1,11 V | 1,15 V |
|  |  | 293 K | 0,83 V | 1,54 V |

### 4.2. Vue d'ensemble

- <u>Système conventionnel</u> - Le pont de diodes fonctionne à température ambiante (Fig. 9). Pour assurer un refroidissement efficace, un dissipateur thermique RS PRO 150×300×83 mm en aluminium anodisé est sélectionné. Une feuille de béryllium et de la pâte thermique sont utilisés pour réduire la résistance thermique tout en garantissant l'isolation électrique entre le boîtier et le dissipateur. A température ambiante, le pont de diodes supporte un courant de 80 A DC en régime permanent dans la plage de température fixée.

- <u>Système étudié</u> - Le pont de diodes fonctionne à température cryogénique (Fig. 10). Pour notre prototype, dans un premier temps, nous avons conservé le dissipateur thermique, bien qu'il serait possible de fortement diminuer sa taille [4]. L'ensemble est refroidi à 77 K dans un bain ouvert d'azote liquide.

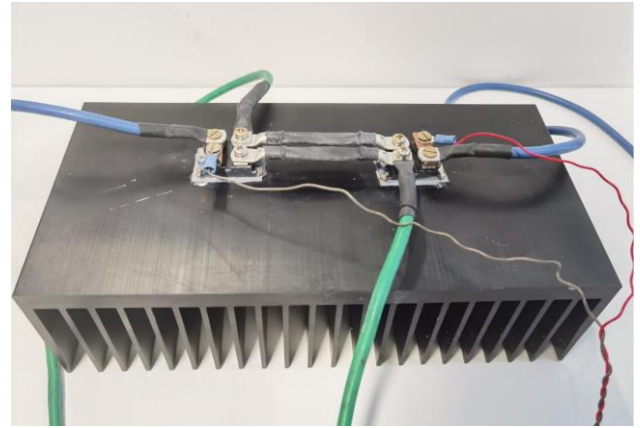

Fig. 9 - Prototype de pont de diodes.

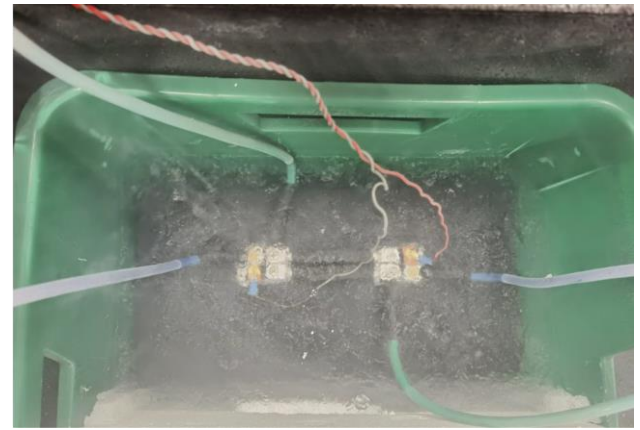

Fig. 10 - Test du pont de diodes à température cryogénique.

### 4.3. Résultats expérimentaux

Pour les tests, le côté AC du pont de diodes est connecté à une source de tension alternative de 50 V RMS et le côté DC est connecté à une charge résistive. La Fig. 11 montre les pertes mesurées du pont de diodes $P_D$ en fonction de la puissance côté DC $P_{dc}$ à température ambiante et à température cryogénique. Les pertes diminuent en moyenne de 27,6 % lorsque le pont de diodes fonctionne à 77 K. La Fig. 12 montre que l'efficacité du pont de diodes s'améliore alors de plus de 2% pour une puissance DC de 1 kW.

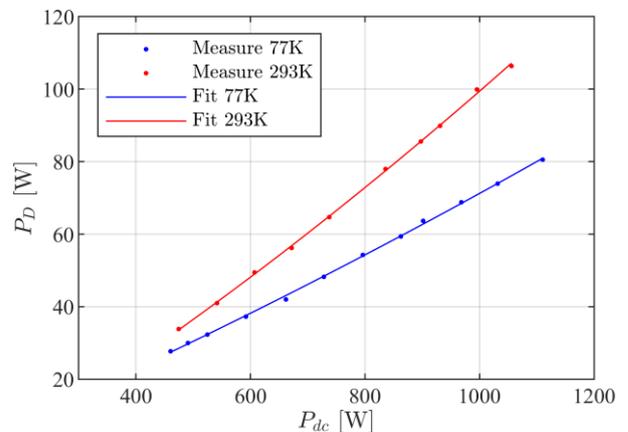

Fig. 11 - Pertes mesurées du pont de diodes, à température ambiante et à température cryogénique ($P_D = P_{ac} - P_{dc}$).

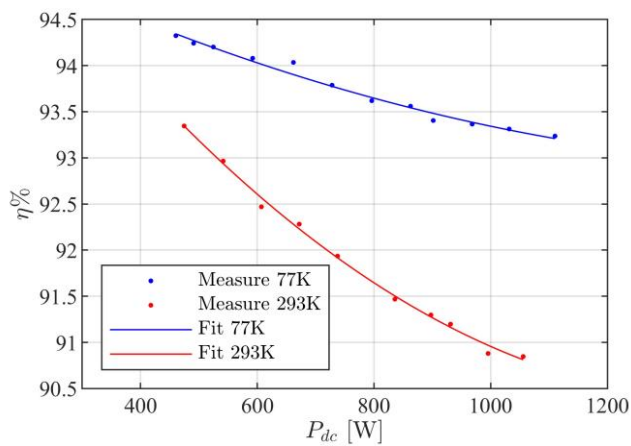

Fig. 12 - Efficacité mesurée du pont de diodes, à température ambiante et à température cryogénique. L'efficacité ne tient pas compte de l'efficacité du système de refroidissement ($\eta = P_{dc}/P_{ac}$).

5. CONCLUSION

Dans cet article, nous avons étudié deux systèmes d'alimentation différents pour les dispositifs supraconducteurs à courant continu. Le premier système, qui est conventionnel, est composé d'un transformateur et d'un pont de diodes fonctionnant à température ambiante, nécessitant des amenées de courant pour passer du milieu à température ambiante au milieu cryogénique. Le deuxième système intègre un transformateur avec un enroulement secondaire supraconducteur, associé à un pont de diodes fonctionnant à température cryogénique, éliminant ainsi le besoin d'amenées de courant. Nous avons tout d'abord présenté qualitativement les avantages et inconvénients des deux systèmes. Nous avons ensuite comparé expérimentalement les performances d'un transformateur conventionnel et de son équivalent supraconducteur. Les résultats ont montré que le prototype de transformateur supraconducteur développé présente une résistance des enroulements et une inductance de fuite réduites par rapport au transformateur conventionnel. Finalement, nous avons mis en évidence que certaines diodes sont adaptées à un fonctionnement à température cryogénique, tandis que d'autres ne conviennent pas. Nous avons confirmé expérimentalement qu'un pont de diodes, utilisant des diodes adaptées, présente des pertes réduites lorsqu'il fonctionne à température cryogénique. L'étape suivante consiste à concevoir et à construire un transformateur supraconducteur adéquat et à le coupler à un pont de diode cryogénique pour alimenter un dispositif supraconducteur sans utiliser d'amenée de courant.

Ces travaux expérimentaux marquent une première étape vers la réalisation d'un système d'alimentation complet pour un dispositif supraconducteur à courant continu. Un tel système pourrait offrir des avantages en termes de réduction de volume et de masse, ainsi qu'une amélioration de l'efficacité globale du système pour certaines applications. Ces recherches ouvrent la voie à des développements futurs dans ce domaine et contribuent à l'avancement de la technologie supraconductrice.